# Trait-based numerical model for mixotrophic phytoplankton and application in Singapore water

My Ha DAO

Center for Environmental Sensing and Modeling (CENSAM), Singapore-MIT Alliance for Research and Technology (SMART).
S16-05-08, 3 Science Drive 2, Singapore 117543

**Abstract**

A numerical model for mixotrophic phytoplankton is described in this paper. In contrast with traditional approach where nutrient uptake rates are constrained by a predefined growth rate, this model uses empirical traits to compute nutrient uptake rates, and then the growth is controlled by the nutrient uptake. Simple but meaningful traits for heterotrophy are derived by analogising heterotrophic mode with phototrophic mode. The trait-based approach could reduce the model parameterization significantly. Model performance evaluation against laboratory experiments of various phytoplankton species has shown remarkable successfulness. Using a single set of model parameterization, the model is able to capture well the growth rate, nutrient consumption, accumulation of none-limiting nutrients, increase of cell size of nutrient-starved cells, surge uptake and rapid population growth of starved cells when nutrient is added as well as the mixotrophic interaction of different pairs of predator-prey. The model for mixotrophic phytoplankton is couped with a hydrodynamic model to simulate algal blooms in Singapore water. The model has shown a great potential for practical applications.

## 1  Introduction

Phytoplanktonic algae exist in marine water even if the water seems clear. But when oceanographic conditions are suitable, these algae can multiply rapidly in few days to become an algal bloom having tens of millions of cells in a litre of sea water. Algal blooms are usually dominated by single species or a species group which have the potential to be harmful to animals and humans. These blooms are called "red tides" or Harmful Algal Bloom (HAB). HABs can also occur in estuarine, and fresh waters.

The harmfulness of a bloom is the negative effect to the ecosystem. A common negative effect of an HAB is the production of toxins that could cause immediate kill of marine animals or being accumulated over the food chain and become dangerous to higher level animals and humans. Another notable negative effect of an HAB or a general algal bloom is the overgrowth at the water surface over a long period that prevents the light to penetrate into water column, shading the bottom vegetation. The death of bottom vegetation and the excessive algal biomass consumes oxygen in the water (dissolved oxygen), leading to oxygen depletion that suffocates marine animals. Other negative effects of algal blooms although not harmful include the odour smell, skin irritations, cost for beach clean-up, etc. Most of algal blooms last several weeks, but the life span of any individual phytoplankton is rarely more than a few days. Can we predict the occurrence and degree of harmfulness of such HABs? It is possible at least for short terms (days to weeks). Because of different effects that an algal bloom can cause to the ecosystem, a predictive model should be able to accurately estimate not only the duration and intensity (biomass) but also the species composition, especially potential toxic species, of the bloom. There are many numerical models that have been developed to forecast HABs, but probably very few of them has gone into very species specific that allows a species to outcompete the others in certain living conditions and vice-versa.

That said, an important goal for developing a process-based algal bloom model is to pick up the emergence of species out of a phytoplankton community. To do so, the model must have a predictive capacity for individual species responding to various known and predicted nutrient conditions (Glibert et al. 2010). In traditional models, phytoplankton is often modelled based on phytoplankton dynamics which are obtained from averaging the properties of a population or a broad group of phytoplankton. The most basic model of this type is in the nutrient-phytoplankton-zooplankton (NPZ) form coupled with a fixed stoichiometry (usually Redfield) and Monod kinetics. NPZ model was successful at capturing bulk system properties such as chlorophyll and net primary production at large space and time scales such as in open water and seasonal scales. This type of model is, however, unsuitable and could be dysfunctional to describe response of algal under variable nutrient conditions which is crucial for modeling algal bloom (Glibert et al. 2010).

When the phytoplankton group shrinks to a particular species, the group properties become species-specific or individual-based. Thus, there is a transition from the population-averaged level to the species-specific level dependent on the definition of "individual". For example, "individual" could be "small diatom" to be separate from "large diatom" in a phytoplankton community or it could be "*Skeletonema costatum*" to be distinct from other small diatoms. At a more detailed level of individual, numerical models are theoretically reproduce more accurately the dynamics of phytoplankton individuals as more species specific properties could be modelled. An excellent model of this type was presented in Flynn & Mitra (2009). This model takes into account complex pathways of nitrogen metabolism, mixotrophy, photoacclimation, kleptochloroplastic photosynthesis, phototrophic-heterotrophic interactions. This type of models is no doubt very advance for studying transient physiological responses of phytoplankton.

There are, however, trade-offs to models of this type for practical applications. The number of model parameters must be fitted often overwhelms the ability to constrain them properly from observations (Denman, 2003). The number of state variables and parameters will increase drastically if the detailed individual-based model is applied to every individual in a phytoplankton community. Furthermore, many characteristics of phytoplankton at the species-specific level are very less understood, which may result in inadequate parameterization and eventually deteriorating the model accuracy (Anderson, 2005). Therefore, there should be an optimum level of complexity of the models (including the species-specific knowledge, the community knowledge and the extent of the phytoplankton community) for practical purposes.

In a recent study, Lichman et al. (2007) has highlighted the roles of functional traits and their trade-offs in structuring phytoplankton communities. Major groups of marine eukaryotic phytoplankton have adopted distinct strategies with associated traits. For example, diatoms have the highest nitrogen uptake, $V_{Nmax}$, and intermediate haft saturation constant for uptake, $K_{SN}$, per carbon quota contrasting with coccolithosphores who have low $V_{Nmax}$ and low $K_{SN}$ while dinoflagellates have widespread of $V_{Nmax}$ and $K_{SN}$. As a result, species of these groups will response differently under different nutrient conditions. Trait types could include life history (reproduction, resting stage), behavioural (motility, mixotrophy), physiological (photosynthesis, nutrient uptake, nitrogen fixation, mixotrophy, toxin production) and morphological (cell size, shape, coloniality) (Lichman & Klausmeier 2008). Trade-offs between traits lead to differentiation of ecological strategies of resource utilization and hence play a key role in determining community structure. Trait-based approach, hence, is an excellent candidate for modelling algal bloom.

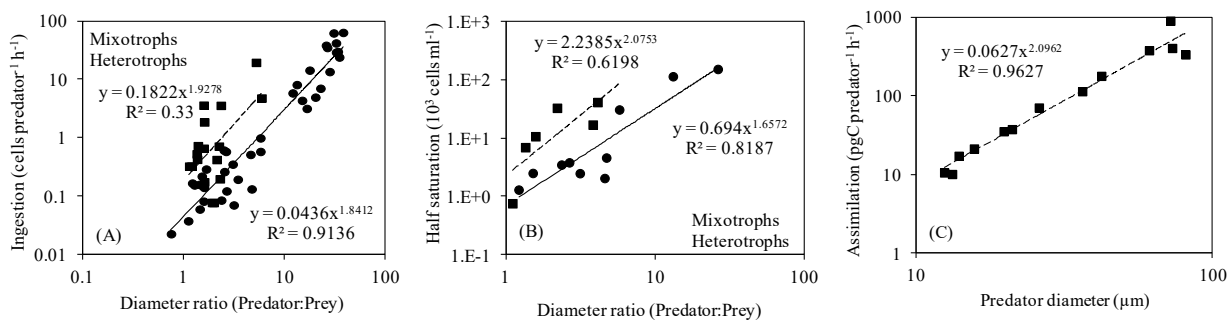

**Fig. 1. Ingestion rate (A), half saturation constant (B) as functions of ratio of predator and prey diameters, and assimilation rate (C) as function of predator size. Results are obtained from analysing data synthesized from literatures (Jeong et al. 2004; 2005a,b,c; 2007; 2010).**

There have been many works on functional traits of phytoplankton. Theoretical studies (Pasciak & Gavis 1974; Aksnes & Egge 1991; Amstrong 2008) hypothesized that maximum nutrient uptake of a phytoplankton cell is a power function of its cell volume. Empirical studies (Smith & Kalff 1982; Litchman et al. 2007; Edwards et al. 2012) reaffirmed the theoretical results and further pointed out that cell size could explain approximately 80-90% the variation in the nutrient uptake of the cell. Interestingly, similar relationships could also be observed in the predator-prey interaction of mixotrophic and heterotrophic plankton although at a lower correlation. That could be explained by analogising prey cells with nutrient ions although the size of a prey cell is relatively much larger than the size of a nutrient ion, and a prey cell could be actively mobile. Nonetheless, power relationships between ingestion rate, half saturation constant and the ratio of predator and prey diameter could be observed (see Fig. 1). Predator's assimilation rate also significantly correlates with the cell size.

These physiological traits are made used to build the core of the phytoplankton model presented in this paper. The physiological trait is expected to pick up advantages of a functional group under a given nutrient scenario. Additional traits including life history, behavioural and morphological are added to the core physiological trait to better model the emergence of individuals. Modelled phytoplankton response under some known scenarios will be tested. Validations of the model to various laboratory experiments will also be presented. The phytoplankton model is then extended to simulate algal blooms in Singapore water.

## 2 The phytoplankton model

Past phytoplankton models are often C-biomass (or N-, P-) based. Growth of a phytoplankton group or community is simulated through the growth of C-biomass. Other nutrients are simulated in terms of ratios to the C-biomass. This approach is well suitable if cell-related behaviours are not considered. However, cell related behaviours were observed to have significant influences on the growth of phytoplankton. A cell changes its size (may doubles its volume) through its life cycle while nutrient uptake rate is almost linearly proportional to the cell volume. Nutrient uptake may also be dependent on the cell shape (Amstrong 2008; Edwards et al. 2012). Surge uptake of a nutrient such as nitrogen or phosphorus may occur when the cell is nutrient-starved (Latasa & Berdalet 1994). Heterotrophic feeding of a predator on a prey could be functions of size ratio, swimming speed of both species, feeding mechanism (Jeong et al. 2004; 2005a,b,c; 2007; 2010) and possibly a random cell-cell contact. Sinking of a phytoplankton cell is dependent not only on its density but also its size and shape following Stokes' law (Waite et al. 1997). Hence, there would be many benefits to operate the phytoplankton mode based on the cell. The C-biomass will not be omitted as it is easily converted from cell population and cell specific C-quota. In other word, the C-biomass is simulated in more details by the cell population and cell specific C-quota. The same could be done for N-, P-biomass.

The numerical presented in this paper is cell-based in which cell population and cell quota are simulated simultaneously. Mathematical formulations of the model will be described in the following sub-sections. Variables and constants used in the formulations are summarised in Table 1 to Table 4.

**Table 1. State variables**

| Parameter | Description | Value/Unit |
|---|---|---|
| $PX$ | Cell nutrient quota, $X \equiv C, N, P, Fe, Si$ | pg cell-1 |
| $PCel$ | Cell concentration | cells L$^{-1}$ |
| $FX$ | Food vacuole content, $X \equiv C, N, P, Fe, Si$ | pg cell$^{-1}$ |

**Table 2. Auxiliary state variables**

| Parameter | Description | Value/Unit |
|---|---|---|
| $PC_{prey}$ | Cell carbon quota of prey | pg cell$^{-1}$ |
| $PX_{child}$ | Cell nutrient quota of children cell | pg cell$^{-1}$ |
| $FC_{prey}$ | Food vacuole content of prey | pg cell$^{-1}$ |
| $PCel_{prey}$ | Cell concentration of prey | cell L$^{-1}$ |

**Table 3. Constants**

| Parameter | Description | Value/Unit |
|---|---|---|
| $a_C$ | Coefficient of C-quota trait function | 0.1874 |
| $b_C$ | Exponent of C-quota trait function | 0.9464 |
| $a_N$ | Coefficient of N-quota trait function | 0.1809 |

| Symbol | Description | Value |
|---|---|---|
| $b_N$ | Exponent of N-quota trait function | 0.9772 |
| $a_P$ | Coefficient of P-quota trait function | 0.0235 |
| $b_P$ | Exponent of P-quota trait function | 1.0185 |
| $a_{Fe}$ | Coefficient of Fe-quota trait function | 0.0002 |
| $b_{Fe}$ | Exponent of Fe-quota trait function | 1.0251 |
| $a_{Si}$ | Coefficient of Si-quota trait function | 1.1515 |
| $b_{Si}$ | Exponent of Si-quota trait function | 0.7422 |
| $a_{Cm}$ | Coefficient of photo C-acquisition trait function | 1.1503 |
| $b_{Cm}$ | Exponent of phototrophic C-acquisition trait function | 0.9549 |
| $a_{Nm}$ | Coefficient of photo N-acquisition trait function | 0.2592 |
| $b_{Nm}$ | Exponent of phototrophic N-acquisition trait function | 0.9311 |
| $a_{Pm}$ | Coefficient of photo P-acquisition trait function | 0.0385 |
| $b_{Pm}$ | Exponent of phototrophic P-acquisition trait function | 1.0271 |
| $a_{Fem}$ | Coefficient of photo Fe-acquisition trait function | 0.0007 |
| $b_{Fem}$ | Exponent of phototrophic Fe-acquisition trait function | 0.8971 |
| $a_{Sim}$ | Coefficient of photo Si-acquisition trait function | 0.4707 |
| $b_{Sim}$ | Exponent of phototrophic Si-acquisition trait function | 0.7694 |
| $rNC_{max}$ | Maximum N:C affecting growth | 0.2 gN gC$^{-1}$ |
| $rNC_{min}$ | Minimum N:C | 0.05 gN gC$^{-1}$ |
| $rPC_{max}$ | Maximum P:C affecting growth | 0.02 gP gC$^{-1}$ |
| $rPC_{min}$ | Minimum P:C | 0.005 gP gC$^{-1}$ |
| $KQ_X$ | Half saturation for quota curve | 10 (N), 0.1 (P) |
| $k_{PAR}$ | Half saturation for light curve | 0.03 (W m$^{-2}$)$^{-1}$ |
| $k_{inh}$ | Inhibition constant for light curve | 0.0001 (W m$^{-2}$)$^{-1}$ |
| $k_0$ | Light attenuation for water | 0.04 m$^{-1}$ |
| $k_{p1}$ | Light attenuation due to live phytoplankton | 4 (µmol P L$^{-1}$ m)$^{-1}$ |
| $k_{p2}$ | Light attenuation due to organic detritus | 2 (µmol P L$^{-1}$ m)$^{-1}$ |
| $Ks_{NO_3}$ | Half saturation for NO$_3$ uptake | 3.6 µM |
| $Ks_{NH_4}$ | Half saturation for NO$_3$ uptake | 3.1 µM |
| $Ks_{PO_4}$ | Half saturation for NO$_3$ uptake | 0.2 µM |
| $Ks_{Fe}$ | Half saturation for NO$_3$ uptake | 0.1 µM |
| $Ks_{SiO_4}$ | Half saturation for NO$_3$ uptake | 3 µM |
| $\sigma_A$ | Inhibition coefficient of NH$_4$ to NO$_3$ uptake | 2.3 µM$^{-1}$ |
| $a_I$ | Coefficient of prey ingestion trait function | 0.042 |
| $b_I$ | Exponent of prey ingestion trait function | 1.851 |
| $a_K$ | Coefficient of half saturation for prey ingestion trait | 0.192×10$^6$ |
| $b_K$ | Exponent of half saturation for prey ingestion trait | 2.383 |
| $a_D$ | Coefficient of digestion trait function | 0.062 |
| $b_D$ | Exponent of digestion trait function | 2.096 |
| $r_F$ | Maximum food vacuole to cell quota ratio | 0.5 |
| $r_{BR}$ | Basal respiration rate | 0.05 gC gC$^{-1}$ day$^{-1}$ |
| $r_{MB}$ | Metabolic respiration rate | 0.005 day$^{-1}$ |
| $r_M$ | Mortality rate | 0.01 day$^{-1}$ |
| $redAA$ | Cost for NO$_3$ reduction and amino acid synthesis | 3.5 gC gN$^{-1}$ day$^{-1}$ |
| $E_A$ | Assimilation efficiency | 0.95 |

**Table 4. Auxiliary model variables**

| Parameter | Description | Value/Unit |
|---|---|---|
| $I_S$ | Irradiance at water surface | W m$^{-2}$ |
| $L_I$ | Light limiting coefficient | |
| $L_{NP}$ | $N, P$ status quotient | |
| $L_Y$ | Nutrient limiting coefficient | |
| $L_{SAV}$ | Shape coefficient | |
| $PX_P$ | Phototrophic nutrient acquisition rate | pg cell$^{-1}$ s$^{-1}$ |
| $PX_H$ | Heterotrophic nutrient acquisition rate | pg cell$^{-1}$ s$^{-1}$ |
| $PX_{Pmax}$ | Maximum phototrophic nutrient acquisition rate | pg cell$^{-1}$ s$^{-1}$ |
| $rXC$ | Ratio of cell nutrient $X$ to carbon quota | gX gC$^{-1}$ |
| $Ig$ | Ingestion rate | preys cell$^{-1}$ s$^{-1}$ |
| $Ig_{max}$ | Maximum ingestion rate | preys cell$^{-1}$ s$^{-1}$ |
| $Ks_{Ig}$ | Half saturation constant for ingestion | cells L$^{-1}$ |
| $IgC$ | Carbon specific ingestion rate | pg cell$^{-1}$ s$^{-1}$ |
| $FC_{max}$ | Maximum food vacuole | pg cell$^{-1}$ |
| $DgX$ | Digestion rate | pg cell$^{-1}$ s$^{-1}$ |
| $POP$ | Phosphorous organic detritus | pg L$^{-1}$ |
| $I_{PAR}$ | Local irradiation | W m$^{-2}$ |
| $D_{pred}$ | Spherical equivalent diameter of predator | μm |
| $D_{prey}$ | Spherical equivalent diameter of prey | μm |
| $r_H$ | Rate of cell lost due to predation | s$^{-1}$ |
| $PV$ | Cell volume | μm$^3$ |
| $Qr$ | Cell division control | g g$^{-1}$ |
| $[Y]$ | Concentration of external nutrients | μM |

## 2.1 Model for growth of the cell specific nutrient quota

Based on phytoplankton cell, changes of the cell specific nutrient quota (or cell nutrient quota) are modelled as

$$\frac{dPX}{dt} = PX_P + PX_H - r_{BR} \cdot PX \qquad (1)$$

where $PX$ represents the cell nutrient quota, $X \equiv C, N, P, Fe, Si$. The subscripts $P, H$ indicate the cell nutrient acquired from phototrophic and heterotrophic nutrient modes, respectively. $r_{BR}$ is the rate of nutrient lost due to cell basal respiration (only for $X \equiv C$), respectively. The initial cell nutrient quotas of non-starved cell could be computed based on the cell volume, $PV$, as

$$PC = a_C \cdot (PV)^{b_C} \qquad (2)$$

$$PX = a_X \cdot (PC)^{b_X} \qquad (3)$$

Here $X \equiv N, P, Fe, Si$. The parameters $a_C, b_C, a_X, b_X$ are constant (see Table 3) and could be dependent on phytoplankton taxa (cyanobacteria, dinoflagellate, diatom, etc).

### 2.1.1 Model for phototrophic nutrient acquisition

Based on physiological traits, the maximum phototrophic acquisition rate of each nutrient of a phytoplankton cell is a function of the cell carbon quota as

$$PX_{Pmax} = a_{Xm} \cdot (PC)^{b_{Xm}} \qquad (4)$$

Here $X \equiv C, N, P, Fe, Si$. The parameters $a_{Xm}, b_{Xm}$ are constant (see Table 3) and could be dependent on phytoplankton type (cyanobacteria, dinoflagellate, diatom, etc) (Litchman et al. 2007). The scale exponents, $b_{Xm}$, of these curves are close to 0.9 (Maranon 2008; Edwards et al. 2012) which indicates that maximum nutrient acquisition rate of a cell is a weakly nonlinear function of cell carbon quota. The maximum carbon phototrophic acquisition rate needs to cover the costs of basal respiration ($r_{BR}$) and of respiration associated with growth at the maximum $N:C$ ($redAA$) (Flynn & Mitra 2009),

$$a_{Cm} = (1 + r_{BR} + redAA \cdot rNC_{max}) \cdot a_{Cm} \qquad (5)$$

The actual phototrophic nutrient acquisition rates of phytoplankton are constrained by the cell nutrient quota ($L_{NP}$), light intensity ($L_I$) and external nutrient concentration ($L_{NO_3}, L_{NH_4}, L_Y$). The formulations are

$$PC_P = PC_{Pmax} \cdot L_{NP} \cdot L_I \cdot L_{SAV} \qquad (6)$$

$$PN_P = PN_{Pmax} \cdot (L_{NO_3} + L_{NH_4}) \cdot L_I \qquad (7)$$

$$PX_P = PX_{Pmax} \cdot L_Y \qquad (8)$$

Here, $X \equiv P, Fe, Si$ and $Y \equiv PO_4, Fe, SiO_4$ accordingly. The parameter $L_{SAV}$ factors the effect of the cell shape on the photosynthesis rate.

Limitation of cell nutrient quota to the phototrophic carbon acquisition rate is calculated using the model of Flynn and Mitra (2009),

$$L_{NP} = \min\{rXCu\}_{X=N,P} \qquad (9)$$

$$rXCu = (rXC \leq rXC_{max}) \cdot \frac{(1 + KQ_X) \cdot (rXC - rXC_{min})}{(rXC - rXC_{min}) + KQ_X \cdot (rXC_{max} - rXC_{min})} \qquad (10)$$

$$+ (rXC \geq rXC_{max})$$

where $rXC$ is the cell specific nutrient-to-carbon quota ratio $X:C, X \equiv N, P$; $rXC_{min}, rXC_{max}$ are the minimum and maximum nutrient to carbon quota ratios, respectively.

Photosynthesis rate is influenced by the light harvesting and utilization of the cell which is often modelled through the light intensity and chlorophyll level of the cell. Photoacclimation is observed in many photosynthetic phytoplankton thus modelling it is useful especially in light-fluctuating regimes (Dubinsky & Stambler 2009). A photoacclimation model (such as Flynn, 2001) may require accurate estimates of the chlorophyll-to-carbon quota ratio ($ChlC$), the initial Chl-specific slope and the photo flux density. The $ChlC$ is generally of order $10^{-2}$ which could be too small for an accurate estimate. Moreover, phytoplankton possesses a diverse set of pigments to capture different parts of the light spectrum. Some species can adjust their pigment composition, just change the Chl-specific slope, to achieve optimum light-harvesting (Stomp et al. 2004; Ting et al. 2002). The photo flux density is also difficult to estimate, especially in real water. The photo flux density in the water would depend on the absorption of the water and the phytoplankton in the water, shading from particles in the water including the phytoplankton itself. A small error in those estimations would deter the effort of modelling photoacclimation. Motive phytoplankton species could also optimize light-harvesting by swimming/floating in response to light (Clegg et al. 2003; Kamykowski et al. 1998; Klemer et al. 1982; Wallace & Hamilton, 1999). These behaviours of phytoplankton are still less understood to be modelled. Hence, in this numerical model, photoacclimation is not considered. The simpler light function in Goebel et al. (2010) is adopted and modified to take into account of the shading effect of the total organic detritus in the water column,

$$L_I = \left(1 - e^{-k_{PAR} \cdot I_{PAR}}\right) \cdot e^{-k_{inh} \cdot I_{PAR}} \quad (11)$$

where $k_{PAR}$ and $k_{inh}$ are the half saturation constant and the inhibition constant for light, respectively. The local photosynthetically active irradiation, $I_{PAR}$, is computed as

$$I_{PAR} = I_S \cdot e^{-k_z} \quad (12)$$

$$k_z = \int_z^0 \left(k_0 + k_{p1} \cdot \sum PP + k_{p2} \cdot POP\right) \cdot dz \quad (13)$$

where $I_S$ is the irradiation at the water surface and $z$ is the water depth; $PP$ and $POP$ are the total phosphorus quota of live phytoplankton cells and the phosphorous organic detritus, respectively; $k_0$, $k_{p1}$, $k_{p2}$ are the light attenuation constants for non-planktonic water, plankton and organic detritus measured in phosphorus, respectively.

Cell shape, in particular the surface area to volume ratio ($SAV$), could also have significant effects on photosynthesis rate (Dao, 2013). Phytoplankton cells having larger $SAV$ could receive more light and nutrients. The effect of $SAV$ could be lumped into the Chl-specific slope in Eq. 20 of Flynn (2001). However, $SAV$ is changing with cell size (linearly proportional to invert of cell radius). As cell size (volume) may be doubled throughout its life cycle, $SAV$ could be decreased by 1.25 (=$1/2^{1/3}$) times or 25%. A coefficient ($L_{SAV}$) that takes into account of the cell shape to the photosynthesis rate of phytoplankton is used in this model. Here, $L_{SAV}$ is computed proportionally to the $SAV$ of the cell.

Constraints to nutrient uptakes due to concentration of external nutrients, $L_Y$, are modelled by the traditional Michealis-Menten approach,

$$L_Y = \frac{[Y]}{Ks_Y + [Y]} \quad (14)$$

where $[Y]$ is the concentration of nutrients $Y$ in the medium surrounding the phytoplankton cell; $Ks_Y$ is the half saturation constant for uptake of the nutrient $Y$. Here, $Y \equiv NH_4, PO_4, Fe, SiO_4$. The half saturation constant for nutrient uptake could be dependent on phytoplankton type (cyanobacteria, dinoflagellate, diatom, etc.) (Litchman et al. 2007). Uptake of $NO_3$ is inhibited by the presence of $NH_4$ in the medium (Goebel et al. 2010),

$$L_{NO_3} = \frac{[NO_3]}{Ks_{NO_3} + [NO_3]} \cdot e^{-\sigma_A \cdot [NH_4]} \tag{15}$$

where $\sigma_A$ is the ammonium inhibition coefficient. The total limitation to the uptake of nitrogen is $L_N = L_{NO_3} + L_{NH_4}$.

There are several complex models for simulating nutrient transport in phytoplankton cells such as those for nitrogen, phosphorus, silicon and iron transports used in Flynn (2001) allowing us to model transient and detailed responses of phytoplankton to the variation of cell nutrient status (Flynn & Mitra 2009). The major trade-off of these models is the parameterization of many species-specific parameters which could prevent them from many practical applications.

### 2.1.2 Model for heterotrophic nutrient acquisition

Mixotrophy is an important nutrient mode of phytoplankton (Burkholder et al. 2008), and in several cases heterotrophic mode of the mixotroph was suspected to be the main cause behind an algal bloom or series of blooms (Jeong et al. 2005b; Mitra & Flynn 2006). Modelling mixotrophy is, however, a challenge in plankton physiology as compared to phototrophy because the intrinsic complexity and lacking quantitative data, especially relating laboratory information to natural field assemblages (Glibert et al. 2010).

In phototrophic mode, nutrient ions may be considered as passive "prey". The behaviour of such prey only follows molecular diffusion rule, and the predator's "capturing" and "ingesting" behaviours may be universal. Concentration of these passive "prey" per "predator" cell is extremely high (even in the nutrient deplete medium), hence a statistical value of "capturing" rate could be used universally and empirical trends are clear for almost every phytoplankton species. In heterotrophic mode, the "prey" are active cells which have various behaviours such as swimming, predation defending. The "predator" also has various predatory behaviours to deal which different preys such as swimming, capturing, ingesting. Moreover, the concentration of preys per predator cell is much lower than that of nutrient ions. Therefore, a statistical value of "capturing" rate may not always work and empirical trends are not clear even for a specific predator-prey pair. Nevertheless, this comparison shows some degrees of analogy between phototrophic and heterotrophic mode that would be useful to build a heterotrophy model.

That said, the ingestion rate ($Ig$) versus prey availability relationship is expected having similar form as the uptake rate versus nutrient concentration, hence Micheallis-Menten function is used to model the prey availability limitation,

$$Ig = Ig_{max} \cdot \frac{PCel_{prey}}{Ks_{Ig} + PCel_{prey}} \tag{16}$$

The maximum ingestion rate, $Ig_{max}$, and the half saturation constant for ingestion, $Ks_{Ig}$, of a predator-prey pair are computed from empirical power functions of the ratio of predator and prey diameters, $D_{pred}/D_{prey}$ (see Fig. 1A, B),

$$Ig_{max} = a_I \cdot (D_{pred}/D_{prey})^{b_I} \tag{17}$$

$$Ks_{Ig} = a_K \cdot (D_{pred}/D_{prey})^{b_K} \tag{18}$$

Parameters $a_I, b_I, a_K, b_K$ are constant (see Table 3) and could be dependent on phytoplankton nutrient type (mixotrophy or purely heterotrophy).

Materials ingested into the predator's food vacuole generally include the prey cells and the food ingested by the preys (if the prey is heterotrophic), $PC_{prey} + FC_{prey}$. The actual ingested material is subjected to the availability of space in the food vacuole which is measured by the balance of the present amount of food in the food vacuole ($FC$) and the maximum food vacuole size ($FC_{max}$). The maximum food vacuole size, $FC_{max}$, is often chosen as a ratio to the actual cell size measured in carbon cell content ($r_F$) and thus grows with the growth of the cell. The formula is:

$$IgC = \min\left[FC_{max} - FC, Ig \cdot (PC_{prey} + FC_{prey})\right] \tag{19}$$

$$FC_{max} = r_F \cdot PC \tag{20}$$

The rate of prey cell lost due to the ingestion of this predator is computed as

$$r_H = IgC/(PC_{prey} + FC_{prey}) \tag{21}$$

If there are multiple preys, operations of Eq. (16)-(21) will be repeated either in random pattern or in sequence of prey preference. Other factors influencing the ingestion rate such as swimming speed difference, feeding mechanism, prey preference (Jeong et al. 2004; 2005a,b,c; 2007; 2010) or anti-grazer adaption could be included but that requires well-formulated traits.

Digestion rate of ingested materials in the food vacuole is dependent on the nutrient demand of the cell, the assimilation efficiency ($E_A$) and must cover a metabolic cost ($r_{MR}$) and a basal respiration rate ($r_{BR}$) (Flynn & Mitra 2009). As an empirical trait (Fig. 1C) is used in this model instead of a maximum growth rate, the nutrient demand is represented in term of cell size (which is consistent with Eq. (4), where cell size defines the nutrient uptake rate). The digestion rates of carbon and other nutrients are modelled as

$$DgC = \frac{1 + r_{MR} + r_{BR}}{E_A} \cdot a_D \cdot (D_{pred})^{b_D} \tag{22}$$

$$DgX = \frac{DgC}{FC} \cdot FX \tag{23}$$

where $FX$ is amount of nutrient $X$ in the food vacuole ($X \equiv N, P, Fe, Si$). Assimilated carbon and other nutrient from the heterotrophic mode is

$$PX_H = E_A \cdot DgX \tag{24}$$

Unassimilated portion of digested materials and all digested $Si$ are excreted as organic matters. The assimilation efficiency for food could decline as food $N:C$ or $P:C$ declines (Flynn & Mitra 2009). Here, given the simplification level of heterotrophic nutrient acquisition model, detailed model for assimilation efficiency may not be necessary.

There were several discussions on how to model the phototrophy-heterotrophy interaction in a phytoplankton. These interactions are often difficult to observed even in laboratory experiment (Flynn & Mitra 2009). Therefore, in this model phototrophy and heterotrophy are modelled additive to the growth of phytoplankton. However, control of heterotrophy over phototrophy could be done through the parameter $r_F$ which could be modelled as a function of nutrient demand (such as availability of light, inorganic nutrient). This function would not be straightforward as, for example, the decrease of light is thought to increase heterotrophy in phytoplankton but some species response in an opposite way (Skovgaard, 1998).

Role of kleptochloroplasts in photosynthesis is clear (Stoecker et al., 1987; Skovgaard,1998). However, modelling of kleptochloroplast is challenging as the knowledge on the processes of sequestering, retention and replacing of chloroplasts from the prey are largely unknown. These processes are even inconsistent among known species. Kleptochloroplast photosynthesis, therefore, is not modelled. This is reasonable since the amount of chloroplast in each prey cell is much less compared to the cell itself (~1%) and given the fixed $ChlC$ model used for photosynthesis in this model.

## 2.2  Model for growth of the cell population

Growth in cell population could be modelled by various approaches (Hellweger & Kianirad, 2007) which include neglecting growth, continuous growth based on external nutrient (Monod kinetics) or internal nutrient (Droop model) and discrete growth based on cell division. In this phytoplankton model, the external and internal nutrients are used in the model for growth of cell nutrient quota as described above. The growth of cell population is modelled by cell division model: when the cell nutrient quotas are enough to create two children cells, cell division takes place and the cell population is doubled. The parameter $Qr$ is used to control the cell division is computed as

$$Qr = \min\left\{\frac{PX}{PX_{child}}\right\}_{X=C,N,P,Fe,Si} \quad (25)$$

where $PX_{child}$ is the cell nutrient quota of the children cell. If $Qr \geq 2$, the cell division takes place and the cell population and cell nutrient quotas are updated,

$$PCel = 2 \cdot PCel \quad (26)$$

$$PX = PX/2 \quad (27)$$

Here, $X \equiv C, N, P, Fe, Si$. Silicon is treated in the same way with the other elements for simplification. A separate model for Si metabolism (Flynn, 2001) or the inhibition of Si to cell division (Brzezinski et al., 1990) in diatom could be used.

The cell division model is useful not only to increase cell population but also to reduce cell size and to redistribute cell nutrient quota, i.e. nutrient quota of the mother cell is distributed into two children cells while keeping nutrient quota ratios unchanged. Cell division model is also useful when simulating excessive growth of nutrient-starved cells after adding the limiting nutrient. For example, cells in phosphorus-limited medium keep taking in carbon, nitrogen and other nutrients and become larger in size (Latasa & Berdalet, 1994; Flynn, 2001). Due to the lack of phosphorus in the cell (not enough to produce two children cells) the cell cannot divide. Once phosphorus is added into the medium, the phosphorus-starved cells uptake phosphorus much faster than cells having the same phosphorus quota but phosphorus-replete due to their larger size [regulates by Eq. (4)]. With other nutrients accumulated in the cells, the cells divide and increase the population rapidly. The cell division redistributes cell contents and size into the new cells, bringing the fast growth rate to a normal rate quickly after 1 or 2 divisions. A feedback function to depress the surge uptake may not be necessary because the redistribution of cell nutrient quota in the cell division model itself produces a feedback mechanism.

Cell loss to mortality and predation of all predators is simply modelled as

$$\frac{dPCel}{dt} = -(r_M + \sum r_H) \cdot PCel \qquad (28)$$

where $r_M$ is the mortality rate. The rate of predation loss, $r_H$, is summed over all predators.

## 3 Model validation against laboratory experiments

Key physiological responses of phytoplankton are used to test the numerical model. These include the steady state and dynamic responses of the cell growth and internal cell quota under various environment conditions. The model is also tested against several laboratory experiments for its ability of capturing responses of phytoplankton in real conditions. In all model tests, the set of constants given in Table 3 is fixed as a default configuration. Details of each test are below.

In the first test, phototrophic steady growth rates of cell population and total cell carbon, nitrogen, and phosphorus (or nutrient specific-biomass) of a phytoplankton in single-algal batch cultures under different light and nutrients (phosphate, nitrate and ammonium) conditions are computed. In each test, one parameter is varied while all others are fixed. The specific growth rates (d$^{-1}$) is computed as $\mu_X = \frac{\log(X^T) - \log(X^0)}{T}$, where $X$ is cell population or the total cell nutrient measured at the initial ($t = 0$) and the end time of the cultures ($t = T$) which is chosen when the cell growth is at stationary phase. The simulated species is a 6200 μm$^3$ dinoflagellate. Results are shown in Fig. 2.

Steady growths of a dinoflagellate under different nutrient modes of pure-phototrophy (with and without presence of a ciliate), pure-heterotrophy and mixotrophy in batch cultures are simulated in the second test. The simulated species are a 10000 μm$^3$ dinoflagellate (predator) and a 2000 μm$^3$ ciliate (prey). In these simulations, inorganic nutrient and light are replete. Steady heterotrophic growth rates of cell population, total cell carbon of the dinoflagellate and prey's cell population (negative growth) at different prey concentrations and predator-prey size ratios are also simulated. Results are shown in Fig. 3 and Fig. 4.

In the third test, dynamic responses of cell population and total cell nutrient (carbon, nitrogen, and phosphorus) of phytoplankton in single-algal batch cultures with a limiting nutrient are simulated. Responses of these parameters of starved cells when the limiting nutrient is added into the depleted media are also tested. The two important limiting nutrients, nitrate and phosphate, are considered. The simulated species is a 6200 μm$^3$ dinoflagellate. Results are compared with phytoplankton grown in nutrient replete cultures in Fig. 5.

In the last test, several laboratory experiments are reproduced by the numerical model. The test includes a batch culture of diatom *Skeletonema costatum* (Takabayashi et al. 2006), batch cultures of dinoflagellate *Prorocentrum donghaiense* under two phosphate-limiting conditions (Ou et al. 2010), and bi-algal batch cultures of dinoflagellate *Dinophysis acuminata* feeding on ciliate *Myrionecta rubra*, and of ciliate *Myrionecta rubra* feeding on crytophyte *Teleaulax sp.* (Park et al. 2006). Growths of cell population in the bi-algal cultures are also compared with that in single-algal cultures of each species. Model results are compared with experiments in Fig. 6 to Fig. 9.

Fig. 2 shows the steady phototrophic growth rates of cell population and total cell nutrient (carbon, nitrogen, and phosphorus) of a phytoplankton in single-algal batch cultures under different conditions of light and nutrient (phosphate, nitrate and ammonium). Under the nutrient replete condition (Fig. 2A), steady state growth of the simulated phytoplankton is saturated at a lower light intensity (~150 μmol photons $m^{-2}$ $s^{-1}$) than the light function itself (~300 μmol photons $m^{-2}$ $s^{-1}$). This is the result of the light attenuation effect on the photosynthetic rate [Eq. (6)-(8)] in batch cultures. The light attenuation, as modelled by Eq. (12) and Eq. (13), is largely dependent on the total phytoplankton and detritus in the medium and hence dependent on the phytoplankton species. Fig. 2A also shows that the growth of cell population is very close to that of carbon and nitrogen. Total cell phosphorus constantly grows approximately 0.3 $d^{-1}$ higher than that of carbon and nitrogen because of the difference in the uptake rate under attenuated light conditions. In real water, phosphorus uptake was shown much less sensitive to the change of light intensity compared with carbon and nitrogen uptakes. Field experiments showed that the dark phosphorus uptake rate is approximately 1.6 times lower than the light uptake (Duhamel et al., 2012) while carbon and nitrogen uptakes are almost zero in dark and much higher in light condition (Kudela et al., 1997). Because of that reason, the light limiting coefficient, which is currently applied on carbon and nitrogen uptakes, is not applied on the photosynthetic uptake of phosphorus in the model. That results in higher uptake of phosphorus.

In Fig. 2B-D, growths of cell population are close to the growths of total cell nutrient of the limiting nutrient and are lower than the growths of total cell nutrient of the non-limiting nutrient. Carbon accumulated in N-starved cells (Fig. 2B, D) is less than that in P-starved cells (Fig. 2C), resulting in larger cell size in P-starved cells. These are consistent with laboratory observations that starved-cells tend to continue acquiring and accumulating none-limiting nutrients, and that P-starved cells are generally larger than N-starved cells. Mathematically, these phenomena are modelled by the use of nutrient uptake as function of cell carbon quota in Eq. (6)-(8), and the limitation of cell nutrient quota in Eq. (9) and (10) in the model.

Fig. 3 shows that carbon acquisition rate of the dinoflagellate via heterotrophic mode is significantly lower than via phototrophic mode. Carbon acquired via mixotrophy slightly differs from that via phototrophy without the presence of the ciliate (Fig. 3A). The difference is mainly at the early stage of the culture when the prey-predator concentration ratio is still relatively high. With the presence of the ciliate (but not as prey), phototrophic carbon acquisition of the dinoflagellate decreases significantly (Fig. 3B). One could explain this phenomenon by the competition for light and inorganic nutrient of the ciliate. Heterotrophy in the mixotrophic dinoflagellate does not only provide additional nutrient source but also removes the prey, which often grows faster, from competing for common resources. The predator will have more resources for growth once the prey is removed from the medium. In this numerical experiment, the direct benefit from feeding the ciliate is less that the indirect benefit from removing the ciliate from the competition for light and inorganic nutrient.

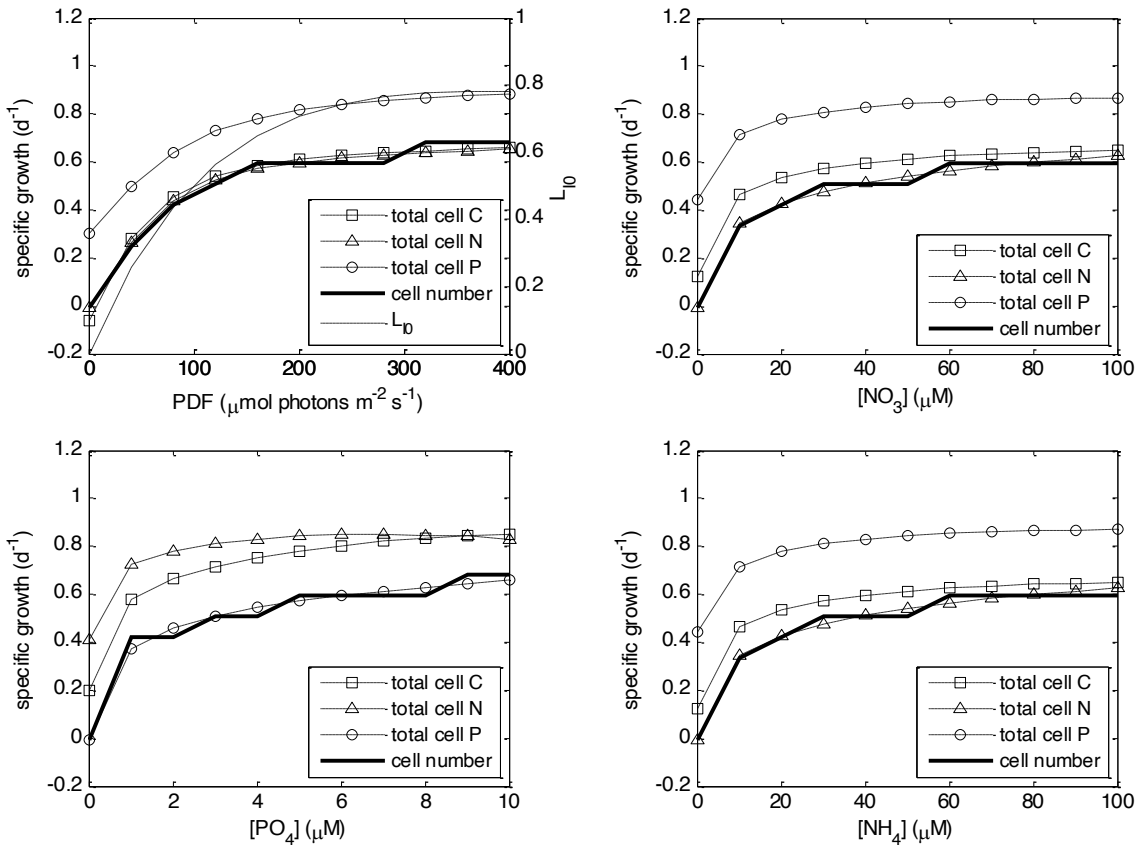

**Fig. 2. Model results of specific growth rates of cell population, total cell carbon, nitrogen and phosphorus at different environment conditions. (A) different photon flux, nutrient replete; (B-D) different nitrate, phosphate and ammonium, other nutrient and light replete; In A-C, ammonium concentration is zero while in (D), nitrate concentration is zero.**

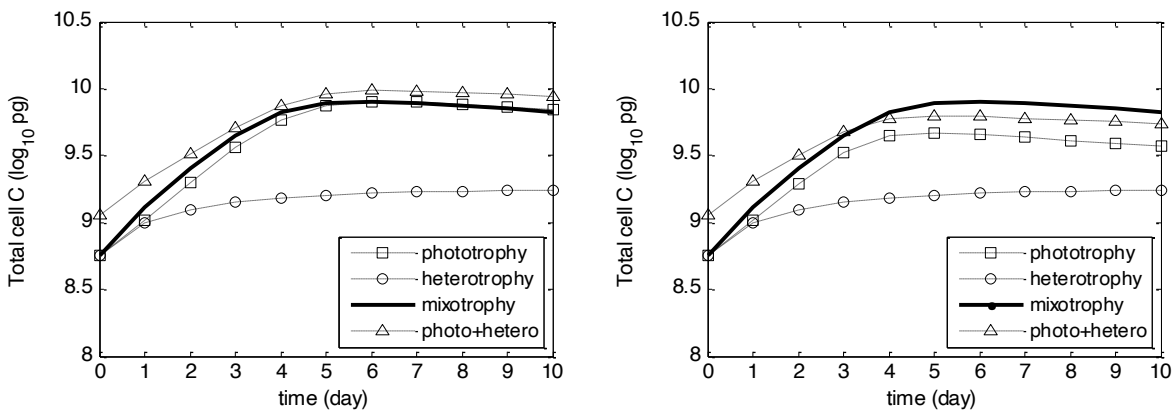

**Fig. 3. Model results of total cell carbon under different nutrient modes: pure phototrophy (square), pure heterotrophy (circle) and mixotrophy (solid line). The lines with triangles are linear summation of phototrophy and heterotrophy lines. (A) pure phototrophy mode of single species, (B) pure phototrophy mode with presence of another species. Inorganic nutrient is replete.**

Model results of pure heterotrophic mode show that the size ratio has more significant effect on the growth of the predator and the removal of the prey than the size of individuals (see Fig. 4, A vs. C and B vs. D). This is due to the maximum ingestion rate and half saturation constant for ingestion being modelled as functions of size ratio. Mathematically, when the diameter ratio doubles (1.5 to 3), the maximum ingestion rate may increase by 4 times (according to the scaling exponent in Fig. 1A and Table 3) but the volume ratio (and hence relative ingested nutrients) reduces by 8 times, resulting in the maximum net ingested nutrient reduced by half. That explains the growth of the predator in the 1.5-diameter-ratio simulations being approximately double of that in the 3-diameter-ratio simulations. However, the use of assimilation rate as function of predator size results in small differences in the growth of the predators of different sizes (Fig. 4, A vs. B and C vs. D). Growth of predator starts at a lower prey concentration if the prey is larger. Minimum prey concentrations for the predator to grow are $10^5$-$10^6$ cells L$^{-1}$. The maximum heterotrophic growth is at prey concentration of $10^9$ cells L$^{-1}$.

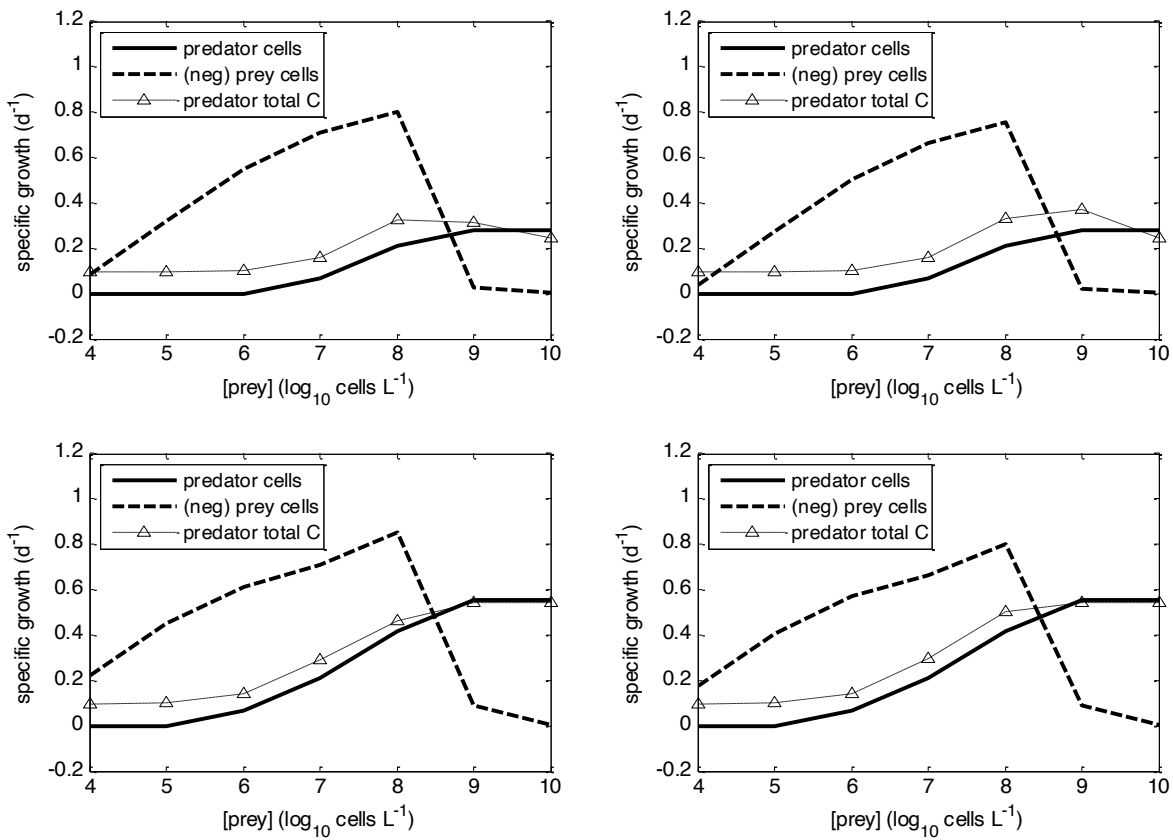

**Fig. 4. Model results of pure heterotrophic specific growth rates of predator cell population and total cell carbon, and negative growth of prey cell population at different prey concentrations. (A, B)** $D_{pred}$:$D_{prey}$ = 3; **(C, D)** $D_{pred}$:$D_{prey}$ = 1.5; **(A, C)** 10,000 μm³ predator; **(B, D)** 5,000 μm³ predator.

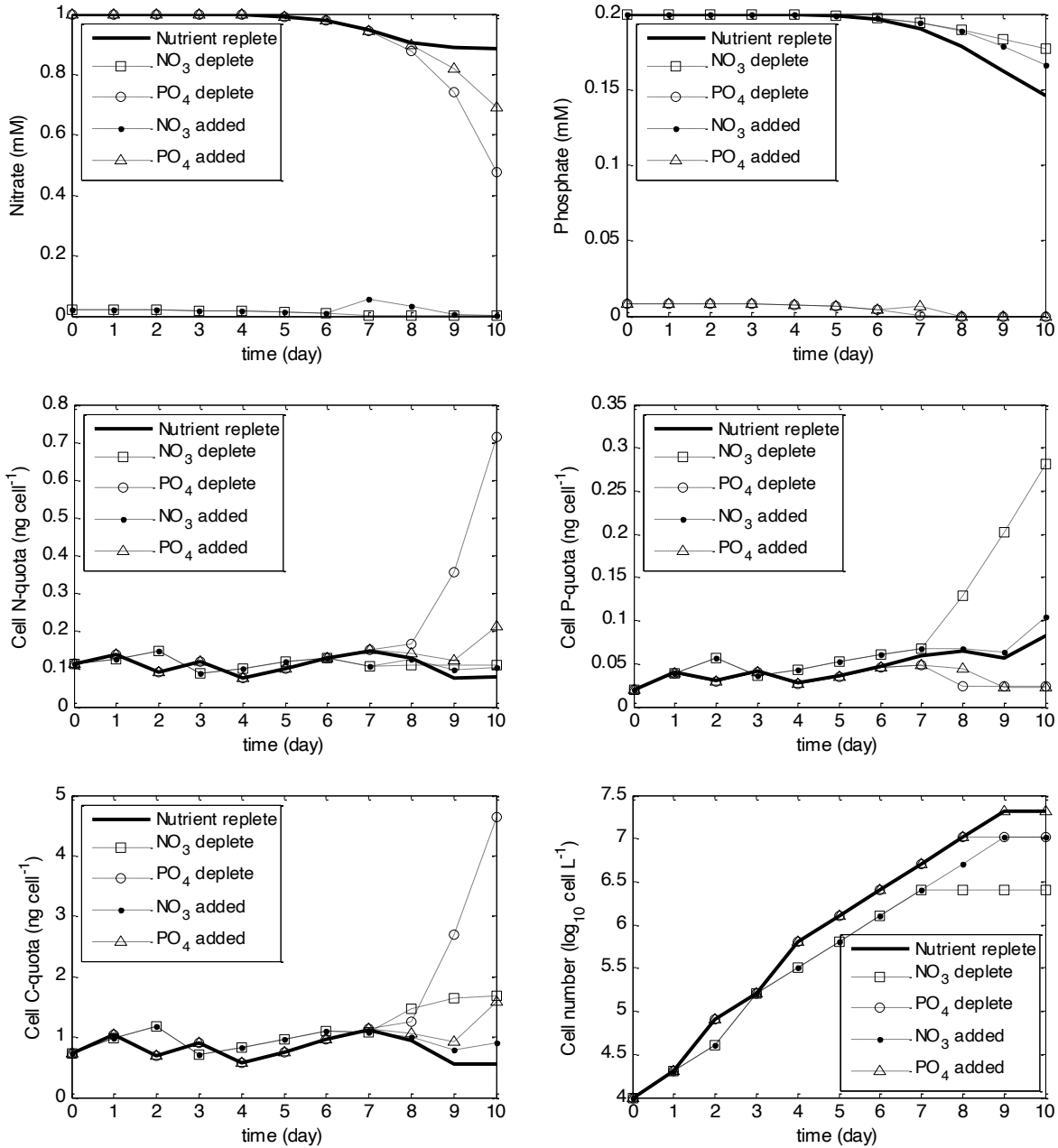

**Fig. 5. Model results of medium nitrate, phosphate, the cell quota of C, N, P and cell population at different nutrient condition: (1) nutrient replete, (2) nitrate deplete, (3) phosphate deplete, (4) nitrate deplete with added nitrate at day 7 and (5) phosphate deplete with added phosphate at day 7. In case (2) and (3) limited nutrient is emptied at day 6.**

Results in Fig. 5 show the responses of cell concentration and cell nutrient status at different external nitrate and phosphate conditions. Nitrate remains replete in the medium in the scenarios 1, 3 and 5 while it is limited in scenario 2 and 4 (Fig. 5A). Phosphate remains replete in the medium in the scenarios 1, 2 and 4 while it is limited in scenario 3 and 5 (Fig. 5B). In scenario 1 where all nutrients are replete, cell quota of N, P and C fluctuate about their initial values. In the nitrate deplete scenario (#2), uptake of nitrate is limited at day 7 while the cell still uptakes phosphorus and some amount of carbon. That results in significant accumulation of phosphorus and carbon after day 7 (square symbol curves). Growth of cell population is significantly lower than that in scenario 1 (see Fig. 5F). Nitrate, when added into the medium at day 7, results in drops in cell quotas of phosphorus and carbon as the larger-than-normal N-starved cell quickly uptakes nitrate and divides, redistributing all accumulated cell nutrient quota (dot symbol curves). As a result, the growth rate of cell population increases rapidly. Similar phenomena could be observed in the phosphate-deplete and phosphate-added scenarios (#3 and 5). However, as shown in Fig. 5E, P-starved cell accumulates more carbon than N-starve cell, resulting in larger cell size. That is modelled by the limitation to carbon uptake by the ratios of cell nitrogen and phosphorus quotas to cell carbon quota [Eq. (9) and Eq. (10)]. It is remarked that some part of growth curves in Fig. 5F are identical. That is because of the cell division model in which cell population at the end of a period will be equal if the number of cell divisions in that period of time is equal, regardless when the cell divisions take place. The difference among these simulations are at the total cell carbon or total cell nutrient which are the product of cell population and cell quota.

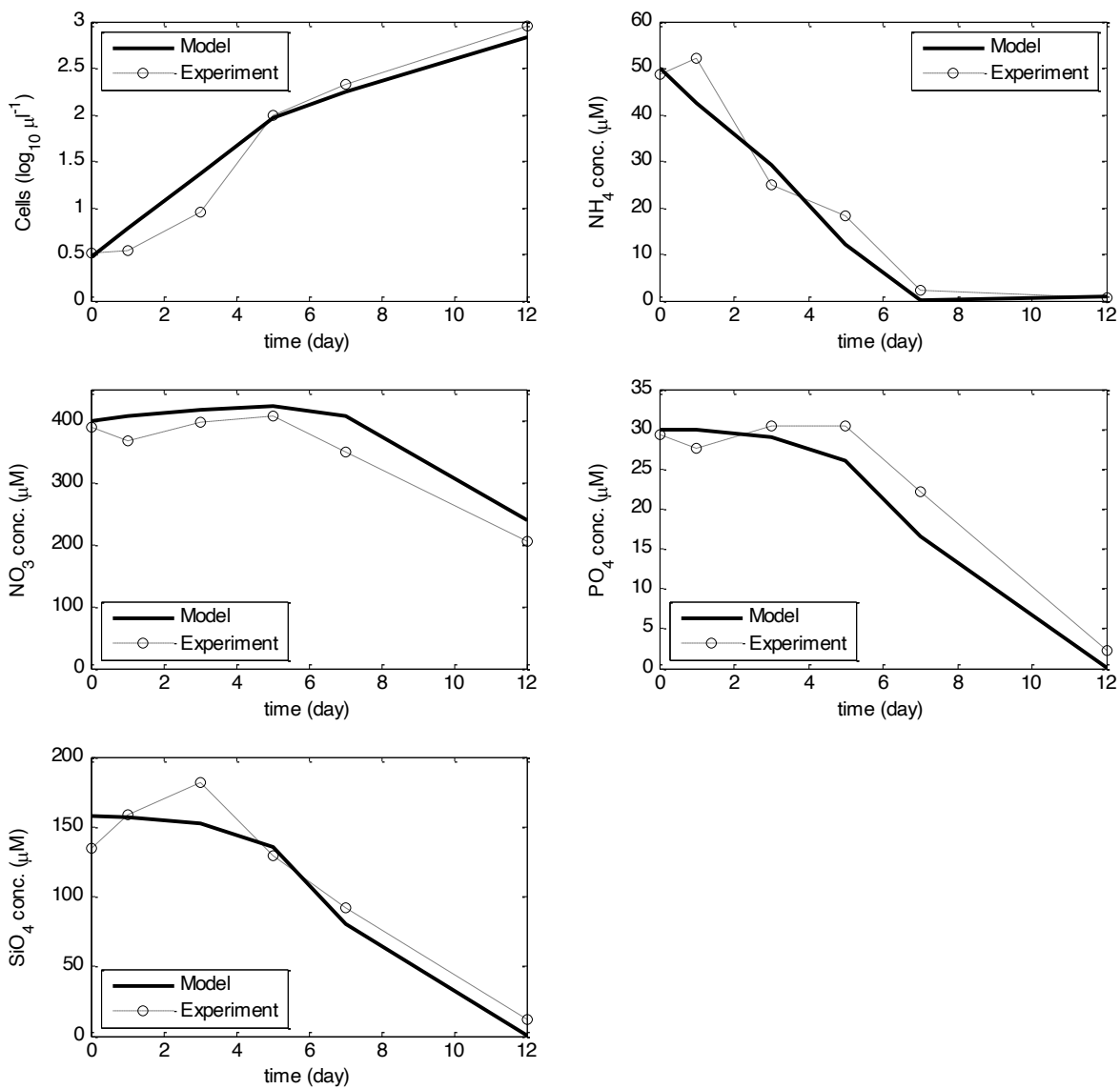

**Fig. 6. Model (solid line) and experimental (circle) results of the cell population, medium ammonium nitrate, phosphate, and silicate from a *Skeletonema costatum* batch culture.**

Results of the numerical simulations in comparison with laboratory experiments are shown in Fig. 6 to Fig. 9. Model results of cell population growth and nutrient consumption of *S. costatum* in Fig. 6 agree very well with experiment results. The initial slow growth of cell population and increase of nutrient in the medium, which were not discussed in the experiment, could possibly due to abnormal mortality and remineralisation of dead cells. In the model, remineralisation process is included but mortality rate is unchanged. Agreements of ammonium and nitrate concentrations between the model and experiment indicate that the simple nitrate-ammonium uptake model of Goebel et al. (2010) is working. Result of the silicon uptake model also matches with the experiment.

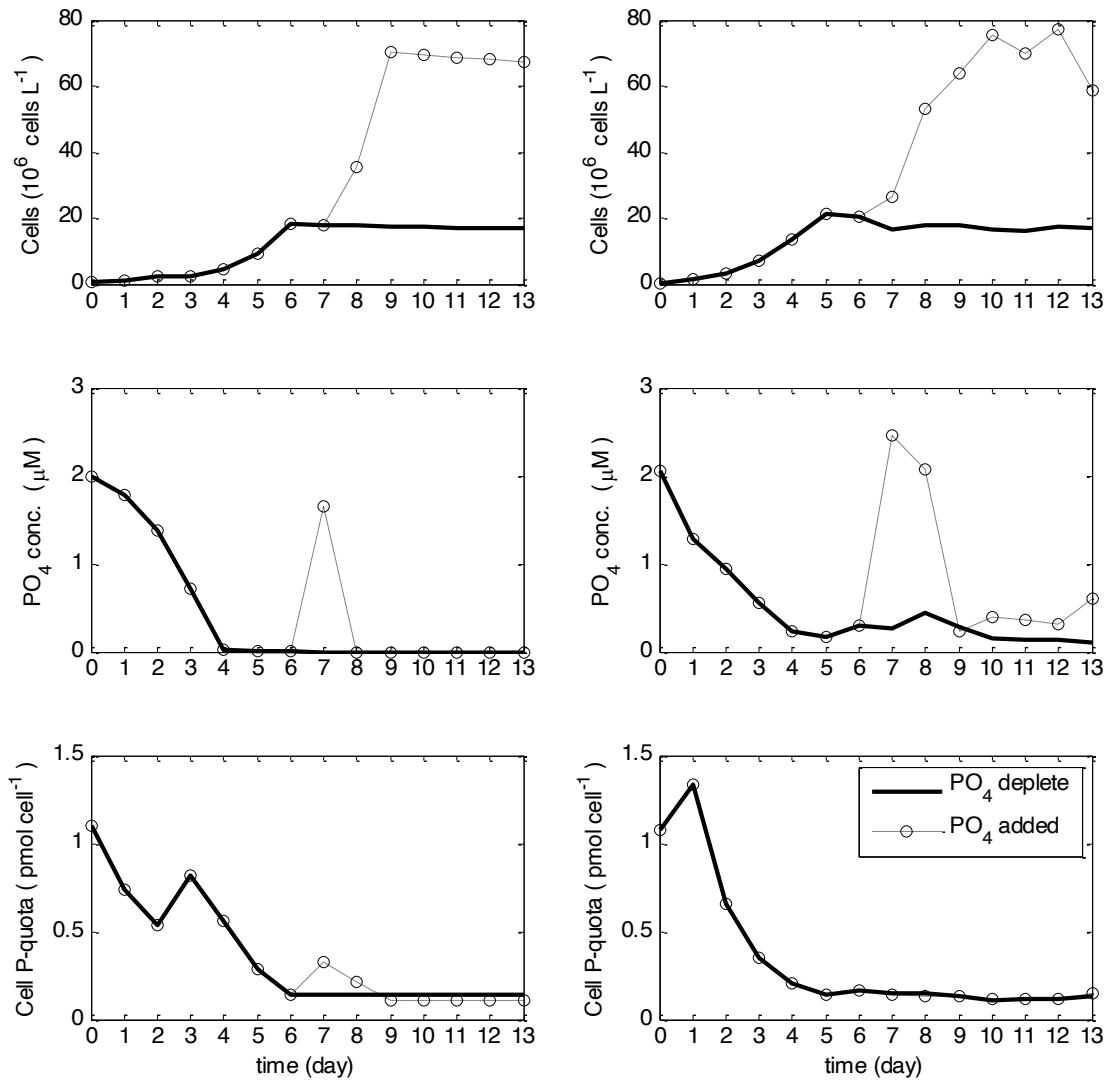

**Fig. 7. Model (left column) and experiment (right column) results of cell population, medium phosphate, and cell phosphorus quota from two cultures of *Prorocentrum donghaiense* under different phosphate-limiting conditions: (1) phosphate deplete and (2) phosphate deplete with added phosphate at day 6.**

Model results of *P. donghaiense* in P-limited medium agree fairy well with the experiment (Fig. 7). Phosphorus cell quota starts at an high level and graduate to a normal level after 5-6 days. The fluctuation in the phosphorus cell quota shown in the experiment is reproduced in the model but at different timing. The cell population growth and phosphate consumption are captured well except for a 1-day difference in the timing of peak cell population. When phosphate is added into the medium containing P-starved cells at day 6, the abnormally rapid growth of cell population and phosphate consumption are captured very well in the model. The peak cell population is close to that obtained from the experiment. The drop in cell population at the last two days of the experiment is, however, unclear and is not present in the model result.

Fig. 8 and Fig. 9 show the model and experiment results of phototrophic and mixotrophic growths of two different pair of phytoplankton species. Phototrophic growths of *D. acuminata* (12,000 μm$^3$), *M. rubra* (6,000 μm$^3$) and *Teleaulax sp.* (100 μm$^3$) are fairly well matched with the experiments (Fig. 8A, B and Fig. 9A, B). Among three species, smaller *M. rubra* grows significantly faster than the largest *D. acuminata* while the smallest *Teleaulax sp.* grows faster than *M. rubra*. Model result of the mixotrophy of the first pair agrees quite well with the experiment (Fig. 8C). In the bi-algal culture, *M. rubra* population slowly grows in the first two days before being rapidly cleared out to almost zero by the predator. *D. acuminata* with the prey grows slightly faster than itself in the single-algal culture. Significant increase of the growth rate of *D. acuminata* could be seen at the first two days when the prey population is still relatively high. Model result of mixotrophy of the second pair in Fig. 9C is not as good as the first pair but shows some degrees of agreement. Population of *M.rubra* is slightly higher when feeding in *Telaulax sp.* compare with that in pure phototrophy mode while the growth of *Telaulax sp.* population is heavily suppressed right at the begin of the culture. Model result of *Telaulax sp.* population relatively deviates farther from the experiment compared to that of *M.rubra* when they are preys. That is because the very small cell size of *Teleaulax sp.* relatively to its predator makes the heterotrophy model less accurate. In overall, however, the numerical model is able to captures fairly well the population responses of the predator and prey at different predator sizes and cell size ratios.

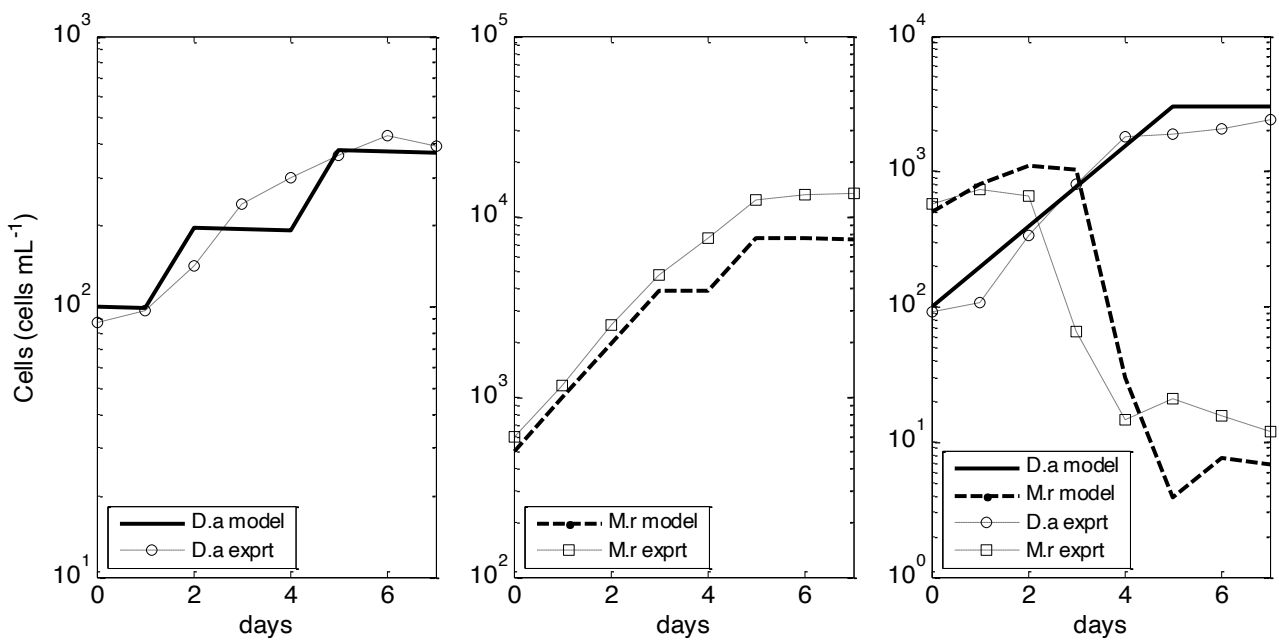

**Fig. 8. Model (thick lines) and experimental results of phototrophic growths of (A) dinoflagellate *Dinophysis acuminata* (D.a), (B) ciliate *Myrionecta rubra* (M.r), and (C) mixotrophic growths of *Dinophysis acuminata* (circles) with *Myrionecta rubra* (squares) as prey.**

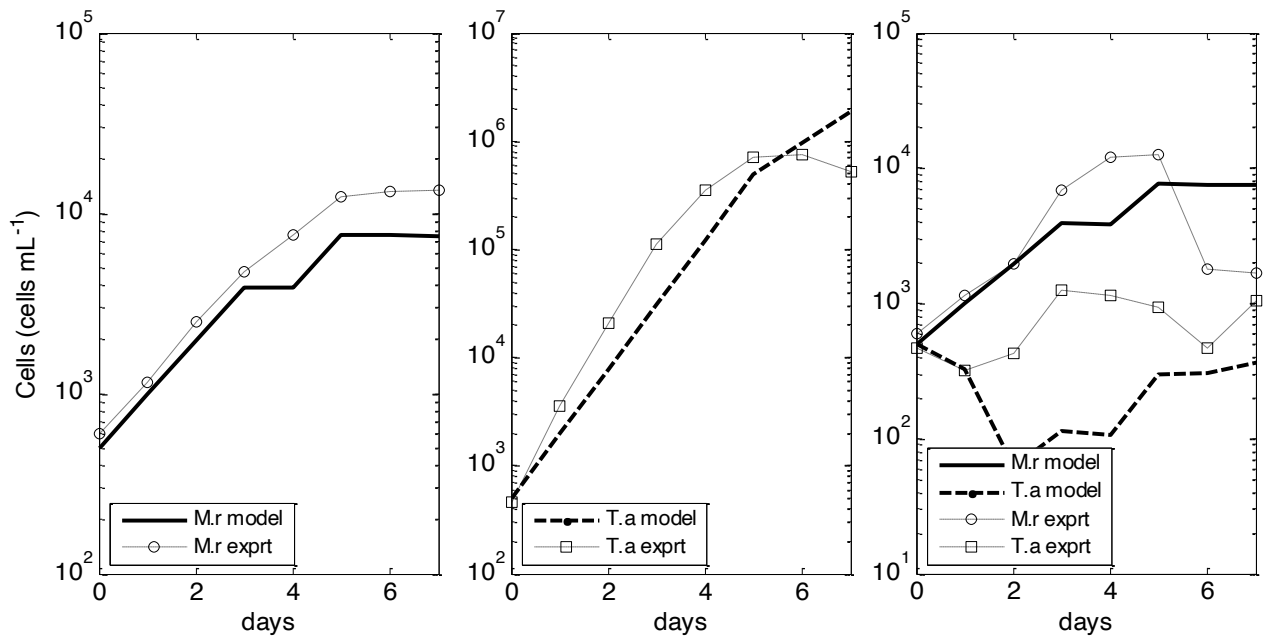

**Fig. 9.** Model (thick lines) and experimental (symbols) results of phototrophic growths of (A) ciliate *Myrionecta rubra* (M.r), (B) cryptophyte *Teleaulax sp.* (T.a), and (C) mixotrophic growths of *Myrionecta rubra* (circles) with *Teleaulax sp.* (squares) as prey.

## 4   Modelling of HAB in Singapore water

The phytoplankton model is extended to simulate algal blooms in Singapore water. There are several possible mechanisms for an algal bloom to occur. Flynn (2001) suggested that an algal bloom model should involve a predator, a HAB species and a non-HAB species. In eutrophic condition, the non-HAB species outgrows HAB species when predator population has not developed yet. When limiting nutrient is deplete and predator has developed, nutrient-stress HAB species produces secondary metabolite (toxic) making it unfavorable (unpalatable) to predator. Predator will eat non-HAB species and themselves. Re-mineralization from predatory activity (and mixotrophy of HAB species) provides additional nutrient for HAB species growth, leading to bloom. Jeong et al. (2004) showed that feed in by larger mixotrophic HAB species on smaller mixotrophic HAB species could be the driving force for the succession of HAB event. In particular, *Cochlodinium* (26 μm) and *P. micans* (27 μm) can feed in *P. minimum* (12 μm) and *P. triestinum* (13 μm), while *P. minimum* and *P. triestinum* can feed in *Amphidinium* (7 μm) and Heterosigma (11 μm). Jeong et al. (2005a,b,c) observed in Korea water that *Amphidinium* and *Heterosigma* bloom followed by *P. minimum* and *P. triestinum*, then by *Cochlodinium* and *P. micans* within 10 days. These observations form the basis for the algal bloom model (AB model) used in this study. Other mechanisms were suggested in Anderson (1998), including cells of HAB species in resting stage (seabed cysts, non-vegetative cells) being geminated and brought to water surface, when environment conditions are favorable (temp, nutrient) or endogenous clock triggered, leading to blooms. However, this mechanism is not considered in the current model.

According to the study of Gin et al. (2000), there are 11 major phytoplankton groups in the Singapore Strait and 8 groups in the Johor Straits identified. Of them, the dominant microplankton (>10 μm) are *Skeletonema* (35% of the total phytoplankton community), *Chaetoceros* (15%), *Eucampia cornuta* (10%), *Eucampia zoodicus* (10%), and *Rhizosolenia stotlerforthii* (5%) in the Singapore Strait and *Chaetoceros* (25%), *Tintinnopsis* (18%), *Skeletonema* (14%) and *Cyclotella stylorum* (5%) in the Eastern Johor Strait. All of them, except the ciliate *Tintinnopsis*, are diatoms. The dominant pico- and nanoplankton (<10 μm) are the cyanobacteria *Synechococcus* and *Trichodesmium*. The microplankton groups contribute about 60% of total chlorophyll in the Singapore Strait and higher portions in the Eastern Johor Strait, which frequently reaches 80-95%. The data also shows that, microplankton totally dominant (>80%) in the sample with high chlorophyll level (>10 μg l$^{-1}$). The dominant species are 10-20 μm and 20-100 μm in size. In term of Chla, diatoms contribute 72% of the in the Singapore Strait and 88% in the Johor Straits. *Synechococcus* contribute another 18% of total Chla in the Singapore Strait but is insignificant in the Johor Strait. *Trichodesmium* was found to blooms in the Johor Straits occasionally during slightly warmer South-West Monsoon (Tham, 1973; Koh, 1998).

Dinoflagellates make up a small portion of phytoplankton community in the Singapore Strait and Johor Strait but several toxic species are found here. Some small unarmoured "gymnodinoid-like" dinoflagellates that have been linked to fish kills in the world were found in Singapore waters (Gin et al. in Wolanski (ed.), 2006). These include species of genera *Karenia* and *Karlodinium*. Another unarmoured *Cochlodinium* dinoflagellate has been linked to fish kill bloom in the Johor Straits in 1987 (Khoo and Wee, 1997). At least four species of golden-brown bi-flagellated raphidophytes have been found in Singapore water. They are *Chattonella marina*, *Chattonella subsalsa*, *Fibrocapsa japonica* and *Heterosigma akashiwo*. These species, except *C. subsalsa*, have also been linked to fish kills in the world. *Dinophysis* species are known to produce diarrhetic shellfish poisoning (DSP) toxins. Among them, *D. caudata* is the most frequent and abundant dinophysoid species although its cell density does not exceed 5 cells l$^{-1}$ (Holmes et al., 1999). The PSP producing dinoflagellate *Gymnodinium catenatum* has been found in Singapore water and its bloom possibly occurred in the past (Khoo and Wee, 1997). Other potential PSP toxin producers are the *Alexandrium* species. Another study (data from CENSAM sea trials, 2011, unpublished) reported the presence of the toxic *A. tamarense* in the Eastern Johor Strait.

**Table 5. Groups of plankton species in EJS for numerical simulation**

| Species | ESD (μm) | Taxa | Toxicity | Group |
|---|---|---|---|---|
| *Trichodesmium erythraeum* | 3 | Cyanobacterium | No | G1 |
| *Synechococcus* | 1 | Cyanobacterium | No | G1 |
| *Alexandrium tamarense* | 22 | Dinoflagellate | Yes | G2 |
| *Chattonella* | 25 | Raphidophyte | Yes | G2 |
| *Heterosigma akashiwo* | 20 | Raphidophyte | Yes | G2 |
| *Chaetoceros* | 10 | Diatom | No | G3 |
| *Skeletonema* | 10 | Diatom | No | G3 |
| *Thalassiosira weissflogii* | 15 | Diatom | No | G3 |
| *Gymnodinium catenatum* | 35 | Dinoflagellate | Yes | G4 |
| *Dinophysis caudata* | 33 | Dinoflagellate | Yes | G4 |
| *Cochlodinium* | 33 | Dinoflagellate | Yes | G4 |
| *Karenia* | 35 | Dinoflagellate | Yes | G4 |
| *Protoperidinium* | 60 | Dinoflagellate | No | G5 |
| *Tintinnopsis* | 60 | Ciliate | No | G5 |

For the convenience of numerical modelling, the plankton in Singapore water are regrouped in 5 groups based on cell size, nutrient mode, taxa, toxicity (see Table 5). The 5 groups are: G1 – Small phototrophs: (cyanobacteria, crytophytes), G2 – Large phototrophs (diatoms, non-diatom phototrophs), G3 – Heterotrophs (large dinoflagellates, ciliates), G4 – Small mixotrophs (dinoflagellates), and G5 – Large mixotrophs (mainly dinoflagellates). General models followed Section 2 are applied for all species, but the kinetics are specific for each group. Bacteria and zooplankton are modelled implicitly by mortality and grazing rates. Possible interactions that might occur in numerical model for EJS include: (i) Under phototrophic mode: G1 population grows fastest due to its small size; G3 population also grows fast but less than G1 and is limiting by silicon; G3 is the main consumer of raw nutrients due to its large size and fast growth; G2 population grows much slower than G3, while G4 and G5 do not grow; light (shading) and oxygen could be inhibition factors; (ii) Under heterotrophic mode: G1 is consumed by G2 and G4, G2 is consumed by G4 and G5, G3 is consumed by G4 and G5, G4 is consumed by G5; (iii) G5 population is controlled by a variable zooplankton grazing and mortality which vary with season and environment.

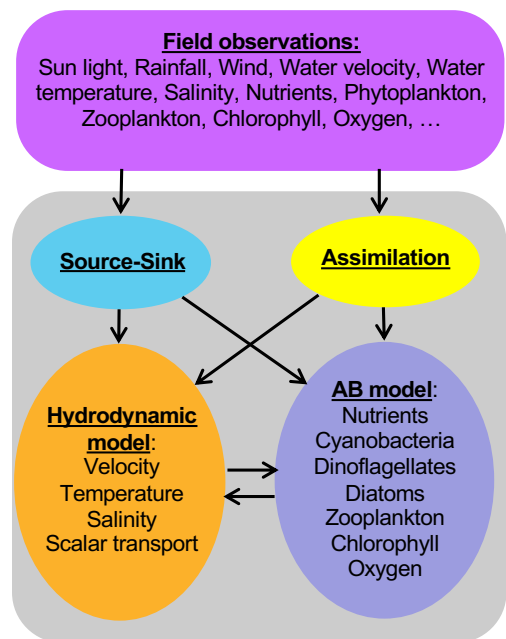

**Fig. 10. Schematic of coupling of hydrodynamic and algal bloom models**

The hydrodynamic in the EJS is simulated by the SELFE model (Zhang and Baptista, 2008). SELFE solves the 3D shallow-water equations, with hydrostatic and Boussinesq approximations, and transport equations for temperature and salinity. These equations are solved by finite-element and finite-volume methods with semi-implicit schemes and Eulerian–Lagrangian method to treat the advection in the momentum equation. Unstructured triangular grids are used in the horizontal direction, while hybrid terrain-following and Cartesian coordinates are used for the vertical direction. SELFE was customized for Singapore Straits and was further zoomed in to the East Johor Strait (Behera and Tkalich, 2014; Xu et al. 2014). The hydrodynamic and algal bloom models are coupled for the simulation of HAB in the EJS (see Fig. 10). The EJS simulation domain is shown in Fig. 11. Boundary conditions of run off, temperature, salinity, concentrations of nutrient, oxygen, plankton cells, and tidal level and velocity are applied at the free surface, river boundaries, and the boundary with the Singapore Strait. Sun light and rain fall are obtained from meteorological measurements. The one-directional arrows indicate river boundaries while the two-directional arrows indicate open boundaries with large water bodies of Johor river and Singapore Strait. Numerical simulations are carried out over a period of at least 7 days in order to capture at least a cycle of algal bloom.

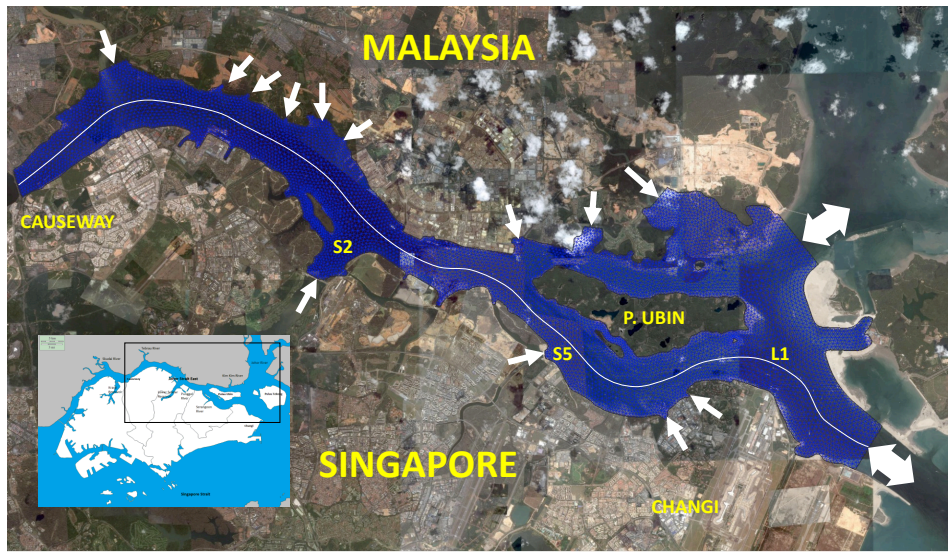

**Fig. 11. Computational domain of East Johor Strait (blue) and boundary conditions (white arrow)**

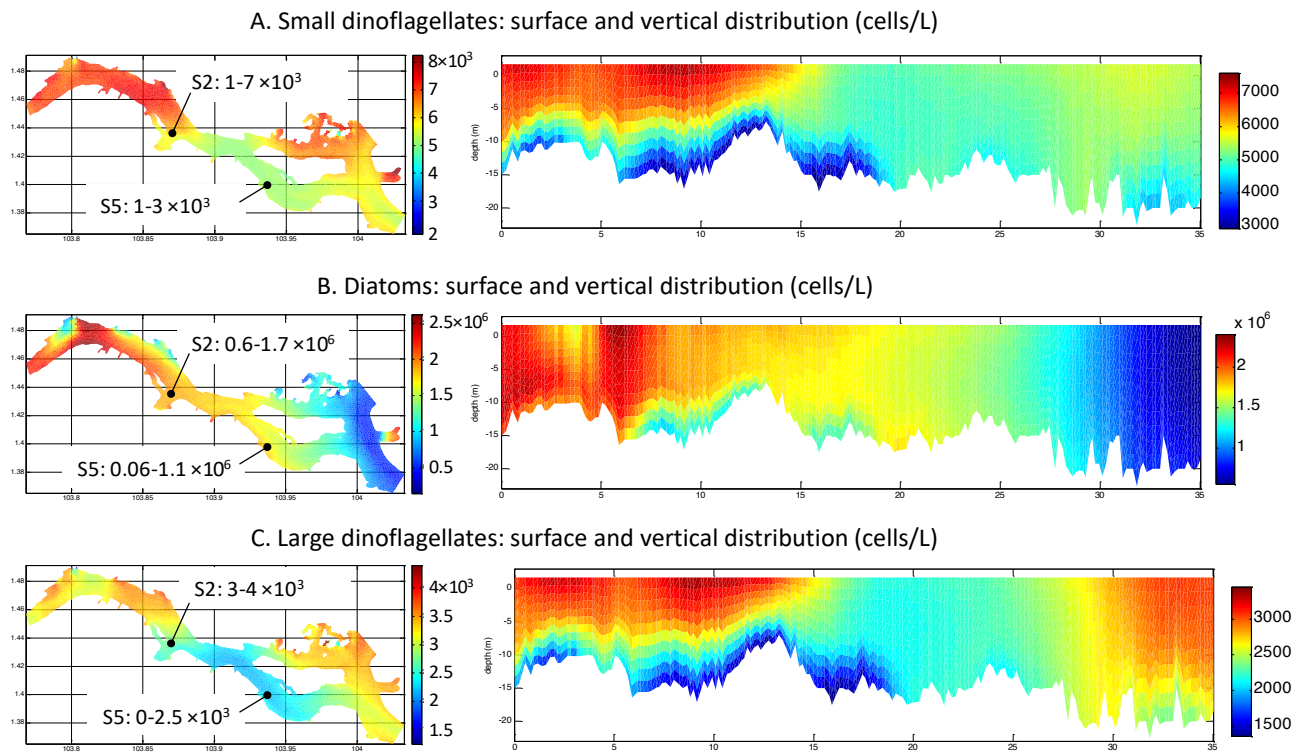

**Fig. 12. Model result of the surface cell concentrations in the EJS (left column) and vertical distributions of cell concentrations along line L1 at a time instant during the neap-tide on July 2012. Field measurements at two locations S2 and S5 indicate the range of cell concentrations in the water column.**

Simulation results of surface cell concentrations of groups G2 (small dinoflagellates), G3 (diatoms) and G4 (large dinoflagellates) in the EJS and vertical distributions of cell concentrations along line L1 at 10 am July 2012 (neap-tide) are shown in Fig. 12. Model results agree quite well with the ranges of cell count at two locations S2 and S5 measured during a sea trials at around same time instant. The model is used to predict concentrations of plankton cells and total chlorophyl during another neap tide period during a North East Monsoon in 2012. Results are shown in Fig. 13.

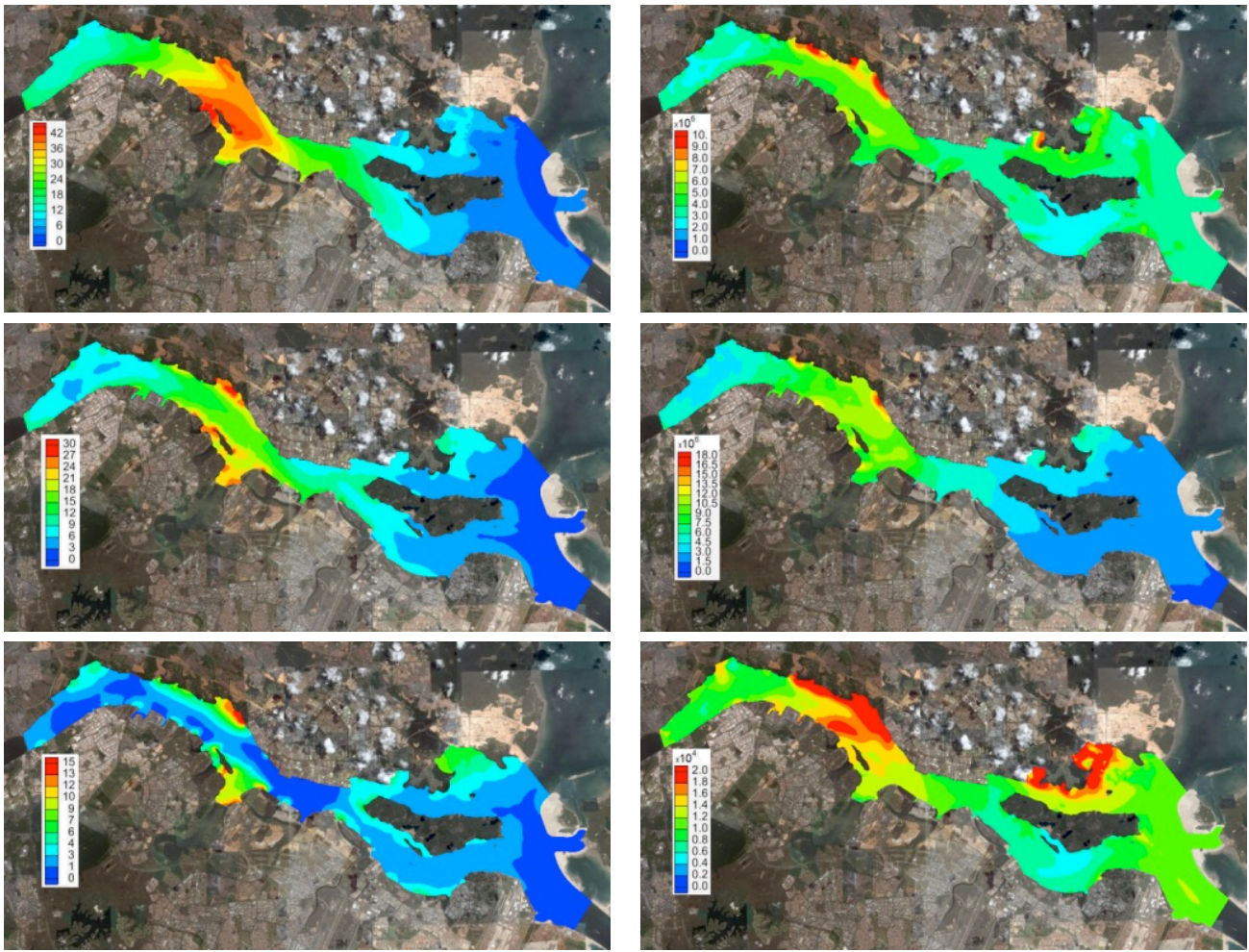

**Fig. 13. Model result of total chlorophyll (left column; from top to bottom: surface, middle and bottom layers; unit: µg L$^{-1}$) and cell concentration (right column; from top to bottom: cyanobacteria, diatom, dinoflagellate; unit: cells L$^{-1}$) in the EJS at 2 pm, 21/12/2012 (neap tide).**

## 5   Conclusion

The paper presents an attempt to construct a numerical model for mixotrophic phytoplankton. The core of the model is the empirical traits that are generalized from experimental data and observations. Important traits such as life history (cell reproduction), physiological (photosynthesis, nutrient uptake, nitrogen fixation, mixotrophy), and morphological (cell size, shape) were employed.

The empirical traits combined with formulations that have been used in traditional mechanistic models provide a good balance of complexities of phytoplankton individual and the extent of phytoplankton community. While some traits, especially those physiological traits of photosynthesis and nutrient uptake, could explain for almost 80-90% of the variability in the community (represented in the dataset), other traits such as those related to taxa and cell shape, and mechanistic formulations could account for the remain 10-20% of small but subtle differentiations among individuals. That not only provides the flexibility for modelling wide range of phytoplankton species with less parameters for parameterization but also allows a narrow group of species to emerge if the environment is at their favourite.

In contrast with traditional approach where nutrient uptake rate is constrained by (or as a function of) a predefined growth rate, the nutrient uptake rate in this model is a function of the cell size and the growth rate is the result of the amount nutrient uptaken. Moreover, the growth of cell population is modelled separately from the growth of cell quota. The latter approach is thus more natural. Numerical results and comparisons with experiments showed that nutrient consumption and population growth could be modelled accurately. The model could also capture well important physiological responses of phytoplankton such as the accumulation of none-limiting nutrients, increase of cell size of nutrient-starved cells, and the surge uptake and rapid population growth of starved cells when nutrient is added. The cell division model is not only to model the growth of cell population and to create a time lag between cell quota growth and population growth, but also play an important role in controlling the nutrient accumulation, surge uptake and rapid growth of cell population.

While phototrophic mode has been well modelled by either trait-based or traditional approaches, the heterotrophic mode is not mathematically well defined. Nevertheless, by using empirical traits as well as mechanistic model of Aksnes & Egge (1991) for phototrophic mode and by analogising heterotrophic mode with phototrophic mode, we were able to derive some simple but meaningful traits for the heterotrophic model. These include the ingestion rate as a function of cell size ratio and the assimilation rate as a function of cell size. Although there remains unresolved complexity in the mixotrophic interaction, the use of these empirical traits in the model has shown remarkable successfulness. Model simulations of mixotrophy presented in this paper agree well with experiments. Numerical simulations also showed that the heterotrophic mode could have little influence on the growth of mixotrophic predator but significantly clear out the prey population. In a plankton community, the mixotrophic predator or non-eatable species would benefit if the prey, which otherwise outgrows them, is removed from the competition for light and nutrient, providing conditions for them to bloom.

Although the numerical model is shown working well for general phototrophs and mixotrophs, several processes could be further enhanced or added into the model to capture species-specific responses. Important processes that could be enhanced include the internal nitrogen metabolism, silicon metabolism, photoacclimation, photo-heterotrophy interaction, prey selectivity, digestion efficiency as well as kleptophotosynthesis. These processes and mathematical formulations have been described in several studies although parameterization is still a challenge (Flynn 2001; Mitra 2006; Flynn & Mitra 2009). Other processes have been shown being crucial for starting or terminating an algal bloom. These include the production of toxins under nutrient stress (Flynn & Mitra 2006;

Varkitzi et al. 2010; He et al. 2010) or under predation stress (Guisande et al. 2002; Sheng et al. 2010; Harvey and Menden-Deuer 2012), the excystment and encystment (Anderson 1998). The model for toxin production under nutrient stress could be easily linked to the cell quota model while that under predation stress could be linked to the ingestion rate of the heterotrophic model. The model for excystment and encystment could be directly linked to the cell population model. With these enhancements, the model is highly suitable for algal bloom modelling.

Nevertheless, the mixotrophic phytoplankton model when coupled with a 3D hydrodynamic model for simulations of HAB in a real world environment has shown excellent results. The HAB model showed a potential of predicting the probability of occurrence, location of an algal bloom and the potential of harmfulness in an areas such as East Johor Strait.